\documentclass[preprint,showpacs,preprintnumbers,amsmath,amssymb,superscriptaddress]{revtex4}

\usepackage{graphicx}% Include figure files
\usepackage{dcolumn}% Align table columns on decimal point
\usepackage{bm}% bold math
\usepackage{color}
\newcommand{\lb}{\left(}
\newcommand{\rb}{\right)}
\newcommand{\ls}{\left[}
\newcommand{\rs}{\right]}

\newcommand{\kap}{\kappa}
\newcommand{\lam}{\lambda}

\newcommand{\vph}{\varphi}

\newcommand{\benu}{\begin{enumerate}}
\newcommand{\eenu}{\end{enumerate}}
\newcommand{\beq}{\begin{equation}}
\newcommand{\eeq}{\end{equation}}
\newcommand{\beqn}{\begin{eqnarray}}
\newcommand{\eeqn}{\end{eqnarray}}

\newcommand{\beqd}{\begin{eqnarray*}}
\newcommand{\eeqd}{\end{eqnarray*}}
\newcommand{\bea}{\begin{array}}
\newcommand{\eea}{\end{array}}
\newcommand{\bcen}{\begin{center}}
\newcommand{\ecen}{\end{center}}
\newcommand{\btab}{\begin{tabular}}
\newcommand{\etab}{\end{tabular}}
\newcommand{\bsub}{\begin{subequations}}
\newcommand{\esub}{\end{subequations}}

\newcommand{\bra}{\langle}
\newcommand{\ket}{\rangle}

\newcommand{\beit}{\begin{itemize}}
\newcommand{\enit}{\end{itemize}}

\def\m@thcombine#1#2{%
  \setbox0=\hbox{$#1$}
  \setbox1=\hbox{$#2$}
  \ifdim\wd0>\wd1
    \setbox0=\hbox to\wd1{\hss\box0\hss}
  \else
    \setbox1=\hbox to\wd0{\hss\box1\hss}
  \fi
  \mathop{\vcenter{
    \offinterlineskip\box0\box1}}}
\def\lesim{\m@thcombine<\sim}
\def\gesim{\m@thcombine>\sim}

\newcommand{\vecr}{\mbox{\boldmath $r$}}
\newcommand{\vecR}{\mbox{\boldmath $R$}}

\begin{document}
%\begin{CJK*}{GBK}{song}
%\preprint{APS/123-QED}

\title{
Asymptotics of neutron Cooper pair in weakly bound nuclei
}

\author{Y. Zhang} \thanks{e-mail: yzhangjcnp@pku.edu.cn}
\author{M. Matsuo}
\affiliation{Graduate School of Science and Technology and Department of Physics,
Faculty of Science, Niigata University, Niigata 950-2181, Japan}
\author{J. Meng}
\affiliation{State Key Laboratory of Nuclear Physics and Technology, School of Physics, Peking University, Beijing 100871, China}
\affiliation{School of Physics and Nuclear Energy Engineering, Beihang University, Beijing 100191, China}
\affiliation{Department of Physics, University of Stellenbosch, Stellenbosch, South Africa}

\begin{abstract}

Asymptotic form of neutron Cooper pair penetrating to the exterior of
nuclear surface is investigated with the Bogoliubov theory for the superfluid Fermions.
Based on a two-particle Schr\"{o}dinger equation governing the Cooper
pair wave function and systematic studies for both weakly bound and
stable nuclei, the Cooper pair is shown to be spatially correlated even in the asymptotic large
distance limit, and the penetration length of the pair condensate
is revealed to be universally governed by the two-neutron separation energy $S_{2n}$
and the di-neutron mass $2m$.

\end{abstract}

% PACS, the Physics and Astronomy
                                 % Classification Scheme.

%\keywords{Suggested keywords}%Use showkeys class option if keyword
                              %display desired
\maketitle

The separation energy of the constituent particle, i.e.,
the minimum energy
needed to remove particle(s) out of a system, influences
strongly surface properties of the system. A characteristic
example in nuclear physics is the neutron halo~\cite{Tanihata,Jonson2004,Tanihata2013},
a dilute
neutron distribution extended far outside the nuclear
surface. This exotic structure
is found near the drip-line, i.e., in the
most neutron-rich isotopes where the neutron separation energy
is reduced by more than one order of magnitude compared with that in naturally abundant nuclei.
It has been considered
that the halo is formed by the last neutrons (the most weakly bound ones) penetrating deeply
into the classically forbidden exterior of the nuclear potential.
However, how they penetrate is a non-trivial question since nucleons are
correlated due to the two-body interaction.
In particular, the neutron pair correlation or the attraction between
the weakly bound neutrons should be taken into account as they
play decisive roles in the formation of
halo~\cite{Hansen87,Bertsch91,Meng-Li,Meng2006,Zhukov93,Jensen2004,
Barranco,Myo2002}.

To answer the question, one may consider
wave function of a "Cooper pair" formed by the last two neutrons.
It can be generally defined by $\Psi_{\rm
pair}(\vecr_1\uparrow,\vecr_2\downarrow)
=\left< \Phi_{N-2} | \psi(\vecr_1 \uparrow)
\psi(\vecr_2 \downarrow) |\Phi_{N}\right>$,
where $\left|\Phi_{N [N-2]}\right>$
is the pair-correlated ground state under interest
with even neutron number $N$ and $N-2$, while $\psi(\vecr_1 \uparrow)$
and $\psi(\vecr_2 \downarrow)$ are neutron annihilation operators
at positions $ \vecr_1$ and $\vecr_2$ with opposite spins.
One needs to know the behaviors
of this Cooper pair wave function in the asymptotic region
$r_1,~r_2\rightarrow \infty$ far outside the nuclear surface.

The correlation of the halo neutrons is often studied for
light-mass two-neutron halo nuclei by means of
the three-body models~\cite{Bertsch91,Zhukov93,Barranco,Hagino05}
and the cluster models~\cite{Kanada-Enyo07,Myo08}, which
suggest that
the two halo neutrons are correlated spatially ---
often referred to as the di-neutron correlation. The asymptotic behavior
is discussed in the Faddeev three-body approach using the
hyperspherical coordinates~\cite{Fedorov1994,Jensen2004}.
However, these models assume a core plus very weakly bound two
neutrons, and the analyses are limited to light-mass drip-line nuclei.

In this Letter, in contrast, we investigate
the asymptotics and the correlation
of the neutron Cooper pair on a more general ground, i.e.,
by using
the selfconsistent mean-field model combined with the
Bogoliubov quasiparticle approach for the pair correlation, which can be applied to essentially
all the self-bound nuclei. In fact,
the Hartree-Fock-Bogoliubov (HFB)
models~\cite{Ring-Schuck,Bender2003} and
the relativistic Hartree-Bogoliubov models~\cite{Vretenar2005,Meng2006}, are successful
in describing not only tightly bound nuclei, but also
neutron-rich nuclei with small separation energies if they are formulated
in the coordinate space~\cite{DobHFB1,Bulgac,Dobaczewski1996,Meng-Li}.
Examples include
two-neutron halo nuclei such as $^{11}$Li~\cite{Meng-Li}
and the giant halo, involving several neutrons, predicted e.g.  in $N>82$ Zr isotopes~\cite{Meng-Zr,Meng2006,GrassoGhalo}.
We note also that the di-neutron correlation in
the Cooper pair wave function is predicted
in the HFB models applied to medium and heavy mass
neutron-rich and stable nuclei~\cite{Matsuo05,Pillet2007} with
separation energies $\sim 2-10$ MeV. On these backgrounds,
we investigate in this study how
the asymptotics of the neutron Cooper pair vary as a function of the
neutron separation energy.

The Bogoliubov's quasiparticle method adopted in the
HFB model is essentially the same as those applied to various Fermion systems
with superfluidity caused by $^{1}S$
short-range attractive
interactions~\cite{Bogoliubov,deGennes,Giorgini}.
The  ground state of a pair correlated nucleus
is approximated as a variational vacuum
$\left|\Phi_0\right>$ of independent quasiparticle states.
The quasiparticles have two-component wave function
$\phi_i(x)=\left[ \varphi_{1,i}(x),
\varphi_{2,i}(x)\right]^T$
with $x=\vecr\sigma$, and obey the HFB equation
\beq
  \left(
    \begin{array}{cc}
      h-\lam & \Delta \\
      -\Delta^* & -h^*+\lam \\
    \end{array}
  \right)
  \left(
     \begin{array}{c}
      \vph_{1,i}(x) \\
      \vph_{2,i}(x)
     \end{array}
   \right) = E_i   \left(
     \begin{array}{c}
      \vph_{1,i}(x) \\
      \vph_{2,i}(x)
     \end{array}
   \right),
 \label{eq:HFB-matrix}
\eeq
known also as the Bogoliubov-de Gennes equation in general~\cite{deGennes}.
The single-particle Hamiltonian, $h=t+U$, includes the kinetic operator $t$ and
the selfconsistent mean field potential $U$.
The Fermi energy is $\lambda$ and the pair potential $\Delta$.
The Cooper pair wave function in this approach is given as
$\Psi_{\rm pair}(x_1,x_2)= \left< \Phi_{0} |
\psi(x_1)\psi(x_2) |\Phi_{0}\right>$, and
 expressed as
\begin{equation}
\Psi_{\rm pair}(x_1,x_2) =
\frac{1}{2}\sum_i
   \vph_{1,i}(x_1)\vph^*_{2,i}(x_2)-(x_1 \leftrightarrow x_2)
\label{eq:Cooper-pair}
\end{equation}
in terms of a sum of the quasiparticle wave functions.

A crucial step to explore the asymptotic form of $\Psi_{\rm pair}(x_1,x_2)$
is to note
that it obeys "two-particle Schr\"{o}dinger equation",
\begin{equation}
 [t(1)+t(2)+v(1,2)]\Psi_{\rm pair}(x_1,x_2)=2\lambda\Psi_{\rm pair}(x_1,x_2)
 \label{eq:cooperpair-wf-eq-fin}
\end{equation}
for $r_1, r_2  \rightarrow \infty$, with
$v(1,2)$ the two-body force between neutrons.

The derivation of this equation is as follows. Operating the single-particle Hamiltonian $h(1)+h(2)$ on $\vph_{1,i}(x_1)\vph^*_{2,i}(x_2)$ in Eq.(\ref{eq:Cooper-pair})
and using Eq.~(\ref{eq:HFB-matrix}), one finds
\begin{eqnarray}
  [h(1)+h(2)-2\lambda]\Psi_{\rm pair}(x_1,x_2)&=& \cr
  &&\hspace{-55mm} -\Delta(1)\sum_i \varphi_{2,i}(x_1)\varphi^*_{2,i}(x_2)
  - \Delta(2)\sum_i\varphi^*_{1,i}(x_2)\varphi_{1,i}(x_1).
  \label{eq:Cooperpair-wf-eq-1}
\end{eqnarray}
Secondly, it can be shown that the r.h.s. of Eq.~(\ref{eq:Cooperpair-wf-eq-1})
will be $-\Delta(x_2,x_1)$
in the asymptotic limit.
Here we use the completeness relation
of the Bogoliubov quasiparticle wave functions,
$
\sum_i \left[ \varphi_{1,i}(x)\varphi^*_{1,i}(x')+
\varphi^*_{2,i}(x)\varphi_{2,i}(x')\right] =\delta_{xx'}
$,
and the known asymptotic behavior~\cite{Bulgac,DobHFB1}
\begin{equation}
\varphi_{1,i}(x) \rightarrow
\left\{
 \begin{array}{ll}
   e^{-\kappa_1 r}/r~ & (E_i< |\lambda|) \\
   e^{\pm i \kappa_1 r}/r~ & (E_i \geq |\lambda|)
 \end{array}
\right., ~\varphi_{2,i}(x) \rightarrow e^{-\kappa_2 r}/r,
 \label{eq:qpwf-asymptotic}
\end{equation}
for $r \rightarrow \infty$ with
\begin{equation}
  \kappa_1 = \sqrt{\frac{2m|E_i+\lam|}{\hbar^2}},~~
 \kappa_2 =\sqrt{\frac{2m|E_i-\lam|}{\hbar^2}},
\label{eq:kap1-kap2-def}
\end{equation}
which leads to  $|\varphi_{1,i}(x)| \gg |\varphi_{2,i}(x)|$.
We can neglect the terms
$\propto \sum_i\varphi_{2,i}(x_1)\varphi^*_{2,i}(x_2)$ in
Eq.(\ref{eq:Cooperpair-wf-eq-1}) for
$r_1,r_2 \rightarrow \infty$. Finally, with the definition
of the pair potential
$\Delta(x_2,x_1) = v(x_1,x_2)\Psi_{\rm pair}(x_1,x_2)$
and asymptotically vanishing potential $U \rightarrow 0$,
we obtain Eq.~(\ref{eq:cooperpair-wf-eq-fin}).

The structure of Eq. (\ref{eq:cooperpair-wf-eq-fin})
is identical to the Schr\"{o}dinger equation for two
interacting particles with the total energy $E=2\lambda<0$. We note that
the two-particle
Schr\"{o}dinger equation, known to hold for the strong coupling limit or the
Bose-Einstein condensate (BEC) regime of the BCS-BEC crossover
phenomenon~\cite{Leggett,Nozieres}, also holds
in the asymptotic limit far outside the surface.
In the following we consider the Cooper pair wave function in the $^{1}S$ channel
($x_1=\vecr_1\uparrow, x_2=\vecr_2\downarrow$).
We also assume the
spherical symmetry of $\left.|\Phi_{0}\right>$.

Adopting the di-neutron coordinate system (the relative coordinate
 $\vecr=\vecr_1 -\vecr_2$ and the c. m. coordinate of the
 di-neutron $\vecR=(\vecr_1 +\vecr_2)/2$), a solution of Eq.~(\ref{eq:cooperpair-wf-eq-fin})
in a separable form can be obtained as,
 \begin{equation}
\Psi_{\rm pair}(\vecr_1,\vecr_2)=
\sum_{L}\int de C^L_e \phi^L_e(r)\Phi_{{\rm d},e}^L(R)P_L(\cos\theta_{Rr}).
\label{eq:Cooperpair-wf-integral}
\end{equation}
This solution is expressed by the relative wave functions %$^{1}S$
$\phi^L_e(r)$ with the angular momentum $L$,
obeying $\ls -\frac{\hbar^2}{2\mu}\Delta_{\vecr}+v(\vecr)\rs \phi^L_e(r)Y_{LM}(\hat{\vecr})
=e\phi^L_e(r)Y_{LM}(\hat{\vecr})$
with the reduced mass $\mu=\frac{1}{2}m$ and the
relative energy $e$, and the c. m. wave function $\Phi^L_{{\rm d},e}(R)Y_{LM}(\hat{\vecR})$ of the di-neutron, behaving at $R \rightarrow \infty$ as
\begin{equation}
\Phi^L_{{\rm d},e}(R) \rightarrow \exp(-\kappa_{{\rm d},e}R)/R
\end{equation}
 with the exponential constant
$\kappa_{{\rm d},e}=\sqrt{2M(2|\lambda| + e)}/\hbar$
for the di-neutron with the mass $M=2m$ and the energy
$E_{\rm d}=-2|\lambda|-e$.

 The $nn$ system has no bound state with $e<0$,
 but in the $^{1}S$ channel it has a virtual state due to
 the large scattering length $a=-18.5$~fm~\cite{Teramond}.
 As a result, at small $r$ and $e \sim 0$
 the $^{1}S$-wave function $\phi^{L=0}_e(r)$ has a large amplitude
 and it depends only very weakly on $e$.
 Provided that $C^{L=0}_0 \neq 0$ and $C^{L=0}_e$
 is smooth as a function of $e$,
 the asymptotic Cooper pair wave function
 (\ref{eq:Cooperpair-wf-integral}) is then
 dominated by the $L=0$ and $e=0$ component.
 Thus we have
\begin{equation}
\Psi_{\rm pair}(\vecr_1,\vecr_2) \rightarrow C^{L=0}_0 \phi^{L=0}_0(r)\exp(-\kappa_{{\rm d},0}R)/R
%\Psi_{\rm pair}(\vecr_1,\vecr_2) \rightarrow C_0 \phi_0(r)\exp(-\kappa_{{\rm d},0}R)/R,
\label{eq:Cooperpair-wf-deriv-2}
\end{equation}
for $R \rightarrow \infty$ and small $r$,
where the exponential constant is
\begin{equation}
 \kappa_{{\rm d},0}=\sqrt{\frac{2M(2|\lambda|)}{\hbar^2} }
 =\sqrt{\frac{8m|\lambda|}{\hbar^2} }.
 \label{eq:kap-cm0}
\end{equation}
The asymptotic form
Eqs.~(\ref{eq:Cooperpair-wf-deriv-2}) and (\ref{eq:kap-cm0}) indicates
the penetration of a di-neutron
correlated spatially at short relative distances, and
its penetration length is controlled only by
di-neutron mass $M=2m$ and
the two-neutron separation energy $S_{2n}=2|\lambda|$.
The amplitudes  $C^{L=0}_{e>0}$
and $C^{L>0}_e$ in Eq.~(\ref{eq:Cooperpair-wf-integral})
influence the behavior at larger $r$, and they may depend on detailed
conditions, e.g. the pair wave function inside
the nuclear surface, the quasiparticle spectra, and the Fermi energy.

In the following, we will examine the asymptotic behavior of the Cooper pair wave
function by performing the selfconsistent HFB calculation with the Skyrme functional
for even-even $^{44-76}$Ca, $^{60-88}$Ni, $^{92-138}$Zr and
$^{120-150}$Sn covering from stable to neutron-rich drip-line nuclei.
The Skyrme parameters are respectively SkM*~\cite{SkMs} for Ca, SLy4~\cite{SLy4} for Ni and Sn,
and SkI4~\cite{SkI4} for Zr.
The pairing force is a density-dependent
contact interaction, with the force strength
$v_0=-458.4$~MeV fm$^{-3}$ and
the energy cut-off $E_{\rm cut}=60$~MeV,which
reproduces the $^1S$ scattering length~\cite{Matsuo2006PRC,Matsuo2010}.
The HFB equation (\ref{eq:HFB-matrix}) is solved
by mesh diagonalization in the radial coordinate space~\cite{Zhangtobe}.
Compared with previous HFB calculations (e.g., Refs.
\cite{Meng-Li,Dobaczewski1996,Matsuo05,Meng-Zr,Meng2006,GrassoGhalo,Zhang2012}),
 we use a larger box size $R_{\rm box}=100$~fm and a
larger angular momentum space $l_{\rm max}=72$ to describe the
asymptotic behaviors of the neutron pairing.

 \begin{figure}%[!h]
\centering
  \includegraphics[width=0.45\textwidth]{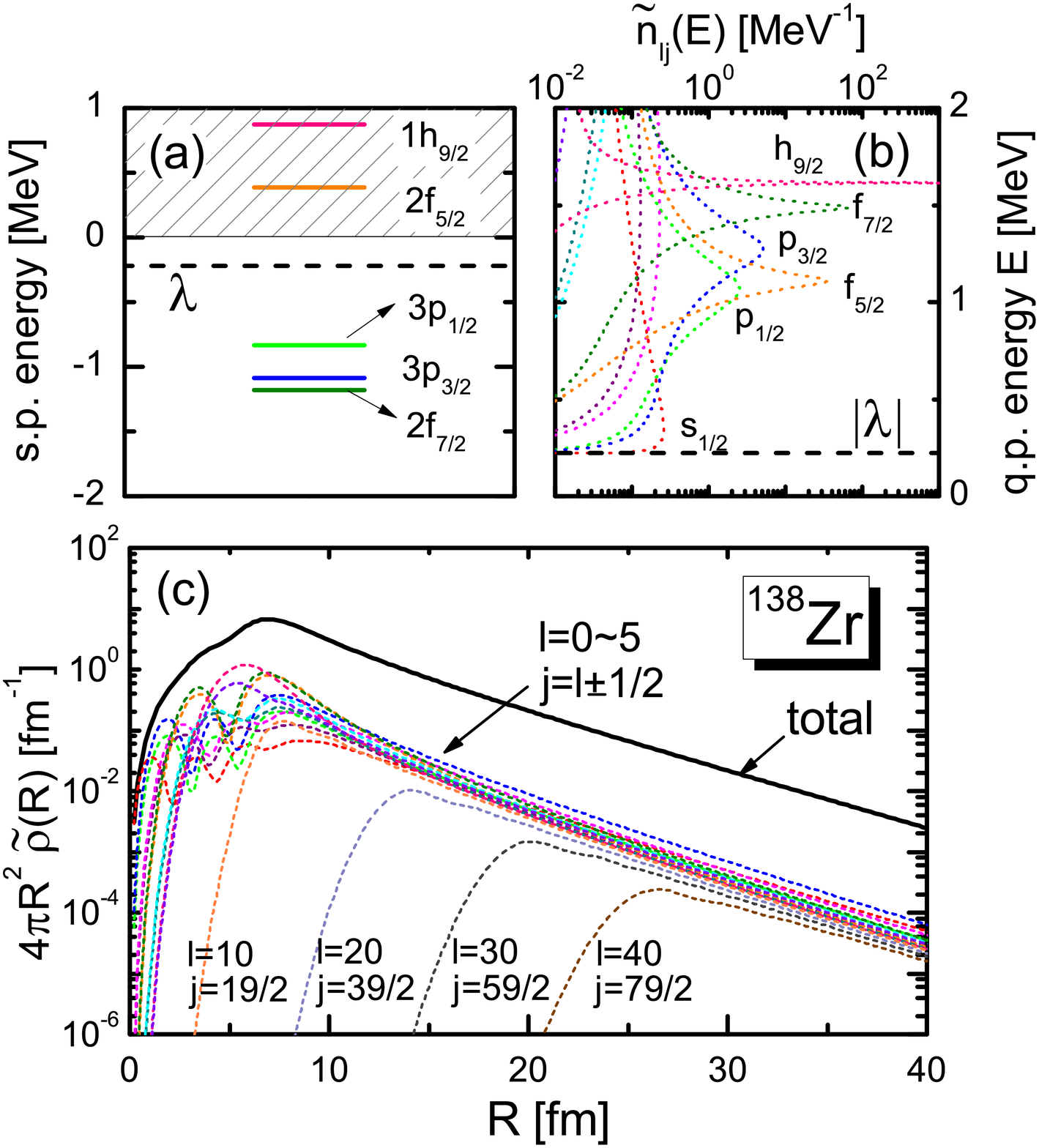}\\
  \caption{(a) Single-particle levels and (b) quasiparticle spectra of neutrons
  in $^{138}$Zr. The quasiparticle spectra are presented in terms of the
  pair number density $\tilde{n}_{lj}(E)$ \cite{Zhang2012}.
  The Fermi energy is denoted by the dashed curve.
  (c) Neutron pair condensate $\tilde{\rho}(R)$ in $^{138}$Zr.
  The solid curve is the total pair condensate
  and the dotted curves are partial contributions $\tilde{\rho}_{lj}(R)$
  from quasiparticle states
  with  $l=0\sim 5$ ($j=l\pm 1/2$),
  and $l=10,~20,~30,~40$ ($j=l-1/2$).
%  The calculated surface radius of this nucleus is $5.48$~fm.
    }
  \label{fig:Zr138-esp-eqp-rhotjl}
\end{figure}

First, let us discuss a very neutron-rich nucleus $^{138}$Zr, which
is predicted to have giant halo structure~\cite{Meng-Zr,GrassoGhalo,Zhang2012}.
Figure \ref{fig:Zr138-esp-eqp-rhotjl} (a) and (b) show
the single-particle levels and quasiparticle spectra of neutrons
in $^{138}$Zr respectively.
The quasiparticle spectra are presented in terms of the
pair number density $\tilde{n}_{lj}(E)$ for
the quasiparticle states with
angular quantum numbers $l=0\sim 5,~j=l\pm 1/2$~\cite{Zhang2012}.
This nucleus has a very shallow Fermi energy $\lambda=-0.22$~MeV
in our calculation, and thus all the quasiparticle
levels turn out to be continuum states
above the threshold $|\lam|$ as seen in Fig.~\ref{fig:Zr138-esp-eqp-rhotjl}~(b).

The neutron pair condensate
(the pair density)
$\tilde{\rho}(\vecR)\equiv \bra \Phi_{0} |
\psi(\vecR\uparrow)\psi(\vecR\downarrow) |\Phi_{0}\ket
=\Psi_{\rm pair}(\vecR,\vecR)$, which is nothing but
the Cooper pair wave function at contact configuration, is shown in
Fig.~\ref{fig:Zr138-esp-eqp-rhotjl}~(c).
It has a huge extended tail with a very gentle exponential slope
due to the shallow Fermi energy.
A large number of partial waves reaching
very high orbital angular momenta
($l\sim 10, 20, 30, 40$ at $R=10,20,30,40$ fm, respectively)  have
coherent contributions of comparable magnitudes to build up the total pair
condensate $\tilde{\rho}(R)$. It is in contrast with the naive
single-particle picture, in which the bound single-particle orbits
located near the Fermi surface, $3p_{1/2},3p_{3/2}$ and $2f_{7/2}$ in the present case,
would be dominant. Moreover it is consistent with the spatially correlated
Cooper pair predicted in our analytic evaluation
Eq.~(\ref{eq:Cooperpair-wf-deriv-2}) since the $l$-coherence up to a large value
$l_{\rm corr}$  implies an angular correlation at  small relative angles
$\lesim \theta_{\rm corr} \sim {\cal O}(1/l_{\rm corr})$ between
two neutrons or equivalently the spatial correlation at
short relative distance $r \lesim R\theta_{\rm corr}\sim {\cal O}(R/l_{\rm corr})$
\cite{Matsuo05}.

\begin{figure}%[!h]
\centering
  \includegraphics[width=0.45\textwidth]{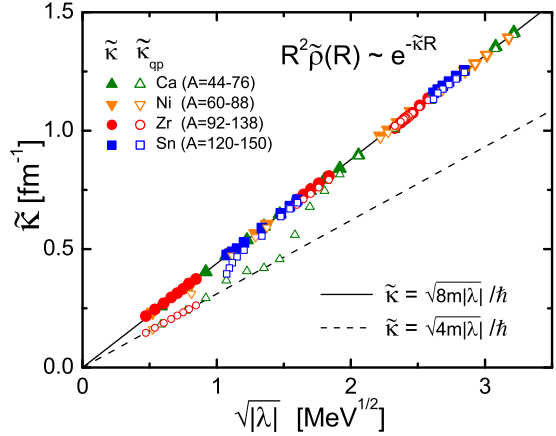}\\
  \caption{The asymptotic exponential constant $\tilde{\kap}$ of the
  neutron pair condensate $\tilde{\rho}(R)$
  obtained from the HFB calculation
  for $^{44-76}$Ca,
  $^{60-88}$Ni, $^{92-138}$Zr, $^{120-150}$Sn,
  plotted with filled symbols as a function of $\sqrt{|\lam|}$,
  where $\lam$ is the Fermi energy.
  The results for $^{48}$Ca, $^{78}$Ni, $^{132}$Sn, $^{110}$Zr and $^{122}$Zr
  are not included due to the absence of pairing.
  The open symbols denote the estimate
  $\tilde{\kap}_{\rm qp}$,
  which is evaluated for the lowest quasiparticle state
  (see text for details).
    }
  \label{fig:kaptVSkapmin-CaNiZrSn}
\end{figure}

A more direct evidence for the analytic expression,
Eqs.~(\ref{eq:Cooperpair-wf-deriv-2}) and (\ref{eq:kap-cm0}),
can be seen in the slope of the exponential tail.
By fitting $R^2\tilde{\rho}(R)=Ae^{-\tilde{\kappa}R}$ at $R=35-40$ fm,
we obtain the exponential constant $\tilde{\kappa}=0.218$ fm$^{-1}$,
which is in good agreement with the analytic value
$\kappa_{{\rm d},0}=0.207$~fm$^{-1}$ calculated with Eq.~(\ref{eq:kap-cm0}).

%##################################################################

The asymptotic exponential constants
$\tilde{\kappa}$ obtained by fitting to the results
for $^{44-76}$Ca, $^{60-88}$Ni, $^{92-138}$Zr,
and $^{120-150}$Sn
are plotted in Fig.~\ref{fig:kaptVSkapmin-CaNiZrSn}. We find that all the
exponential constants $\tilde{\kappa}$ follow quite well the expected
relation $\tilde{\kappa}=\sqrt{8m|\lambda|}/\hbar$
even though the Fermi energy
varies from $\lambda=-10.3$ MeV ($^{44}$Ca) to
$-0.22$ MeV ($^{138}$Zr).

We emphasize that the above results are
non-trivial if one treats the problem from the viewpoint of the
independent quasiparticle basis.  Noting
asymptotic forms of the quasiparticle
wave function, given in Eq.~(\ref{eq:qpwf-asymptotic}),
one may assume that the asymptotics of the pair
condensate $\tilde{\rho}(R) \sim \sum_i \varphi_{1,i}(R)
\varphi_{2,i}(R)$ is dominated by quasiparticle
states with the lowest quasiparticle energy.
This assumption gives an estimate~\cite{Bulgac}
$\tilde{\kappa}_{\rm qp}=\kappa_{1,\rm min} +\kappa_{2,\rm min}$
with $\kappa_1$ and $\kappa_2$
evaluated for the lowest discrete quasiparticle energy $E_{i,\rm min}$
as in Eq.~(\ref{eq:kap1-kap2-def}).
If there is no bound quasiparticle state (the case of shallow Fermi energy),
the lowest quasiparticle state is the one at the threshold
$E_{i,\rm min}=|\lambda|$
for unbound continuum states,
and the estimate would be $\tilde{\kappa}_{\rm qp} =\kappa_{2,{\rm min}}=
\sqrt{4m|\lambda|}/\hbar$~\cite{Bulgac}.
These estimates are also plotted
in Fig.~\ref{fig:kaptVSkapmin-CaNiZrSn}.
In the case of $^{138}$Zr, the above estimate
gives $\tilde{\kappa}_{\rm qp}=\sqrt{4m|\lambda|}/\hbar=0.146$ fm$^{-1}$,
but this is about $30\%$ smaller
than the actual value $\tilde{\kappa}=0.218$~fm$^{-1}\approx \sqrt{8m|\lambda|}/\hbar$.
Clearly a superposition of unbound continuum states is necessary~\cite{Zhang2012}.
We can justify this statement also by noting that the summation in
Eq.~(\ref{eq:Cooper-pair}) over the unbound continuum states
with $E\geq |\lam|$ can be
evaluated approximately as~\cite{Zhangtobe}
\begin{eqnarray}
R^2\tilde{\rho}(R) &\propto& \int_{|\lambda|}^\infty\sin(\kappa_1(E) R)e^{-\kappa_2(E) R} dE \cr
&\sim & K_2\lb \sqrt{\frac{8m|\lambda|}{\hbar^2}}R \rb \sim \exp\lb -\sqrt{\frac{8m|\lambda|}{\hbar^2}}R\rb,
\end{eqnarray}
for $R \rightarrow \infty$. Here $K_2(z)$ is the modified Bessel function.

\begin{figure}%[!h]
\centering
  \includegraphics[width=0.45\textwidth]{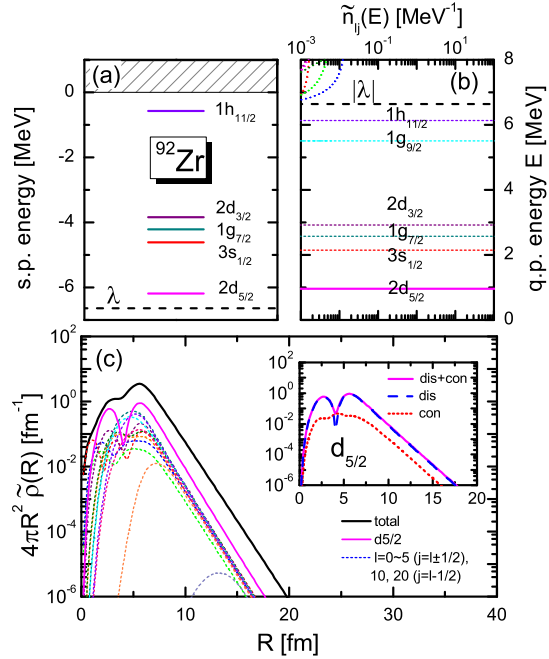}\\
  \caption{Same as Fig.~\ref{fig:Zr138-esp-eqp-rhotjl}
   but for $^{92}$Zr.
  In the inset of panel (c), the partial pair condensate of
  $d_{5/2}$ states (thin solid curve)
  is further separated into those from the discrete $2d_{5/2}$ (dashed curve)
  and continuum states (dotted curve).
    }
  \label{fig:Zr92-esp-eqp-rhotjl}
\end{figure}

Finally we remark that
the microscopic content of the asymptotic Cooper pair
varies with the separation energy
although the asymptotic exponential constant
$\tilde{\kappa}=\sqrt{8m|\lambda|}/\hbar$ is universal
for both neutron-rich and stable nuclei with shallow and deep Fermi energies. An example is
a stable nucleus $^{92}$Zr ($\lambda=-6.6$ MeV).
As shown in Fig.~\ref{fig:Zr92-esp-eqp-rhotjl} (a) (b)
there exist several bound quasiparticle states in this case.
The $2d_{5/2}$ orbit is the one with the lowest quasiparticle energy
that would be occupied by the last two neutrons
in the independent single-particle picture.

The pair condensate $\tilde{\rho}(R)$ and its quasiparticle compositions
$\tilde{\rho}_{lj}(R)$ of $^{92}$Zr are shown in
Fig.~\ref{fig:Zr92-esp-eqp-rhotjl}~(c).
Apart from the steep asymptotic exponential slope
due to the large separation energy,
we find coherent high-$l$ contributions
in the exponential tail, as in the case of $^{138}$Zr.
In addition we see another feature, not seen in $^{138}$Zr, that
there is a significant contribution from the  $d_{5/2}$ states.
It comes from the discrete quasiparticle state
$2d_{5/2}$ as shown in the inset of
Fig.~\ref{fig:Zr92-esp-eqp-rhotjl} (c).
We note that the contribution
of this quasiparticle state,
$\varphi_{1,i}(\vecR\uparrow) \varphi^*_{2,i}(\vecR\downarrow)$
in Eq.(\ref{eq:Cooper-pair}), has an asymptotic exponential constant
$\tilde{\kap}_{\rm qp}=\kappa_{1,\rm min}+\kappa_{2,\rm min} = 1.128$~fm$^{-1}$,
which is almost identical
to the value $\tilde{\kappa}=\sqrt{8m|\lambda|}/\hbar=1.131$~fm$^{-1}$
since $E_i \ll |\lambda|$ (See also the open symbols
lie on the line $\tilde{\kappa}=\sqrt{8m|\lambda|}/\hbar$
for $|\lambda|\gtrsim 4$~MeV in
Fig.~\ref{fig:kaptVSkapmin-CaNiZrSn}).
Thus this quasiparticle contribution remains effectively for physically
relevant range of large $R$.
In this case,
the asymptotic Cooper pair wave function may be generally written as
\begin{eqnarray}
\Psi_{\rm pair}(\vecr_1,\vecr_2) \sim
C'\phi_0(r)e^{-\kappa_{{\rm d},0}R}/R && \cr
&& \hspace{-55mm} +\frac{1}{2}\sum_{i_m}\phantom{}^{'}c'_{i_m}\ls
\varphi_{1,i_m}(\vecr_1\uparrow)\varphi^*_{2,i_m}(\vecr_2\downarrow)-(x_1\leftrightarrow x_2)\rs
\label{eq:Cooperpar-wf-3}
\end{eqnarray}
where the second sum runs over the lowest quasiparticle states.

Equation (\ref{eq:Cooperpar-wf-3}) can cover from stable to drip-line nuclei.
 The second term represents the independent quasiparticle behavior, which survives
if $E_{i,\rm min} \ll |\lambda|$ or $\Delta \ll |\lambda|$, valid for
nuclei close to the stability line~($\Delta$ being the pairing gap).
As the Fermi energy $\lambda$ approaches zero, i.e. in the case of
$E_{i,\rm min} \sim |\lambda|$ or $\Delta \gesim |\lambda|$,
the asymptotic Cooper pair wave function is dominated by the first term,
representing the spatially correlated di-neutron penetration.

In summary, the neutron Cooper pair never
looses its spatial correlation when it penetrates into
the asymptotic region $R\rightarrow \infty$. The
asymptotic  form of the neutron pair condensate is
$\tilde{\rho}(R) \sim e^{-\tilde{\kappa}R}$
with the exponential constant $\tilde{\kappa}=\sqrt{2(2m)S_{2n}}/\hbar$,
which is characterized by the two-neutron separation energy
$S_{2n}=2|\lambda|$ and the di-neutron mass $2m$, irrespective of
whether the Fermi energy  $|\lambda|$ is small or large. It implies
that there is no theoretical upper bound on
the penetration length of the pair condensate, since it
scales as $1/\tilde{\kappa} \propto 1/\sqrt{S_{2n}}$.
The spatial correlation in the asymptotic Cooper pair emerges
explicitly in the weakly bound nuclei satisfying
 $|\lambda| \lesim \Delta$ ($S_{2n} \lesim 2\Delta$), while for
 large separation energies the independent particle features coexist.

In the present analysis we have employed the fact that
 the $^{1}S$ interaction has a large scattering length.
 If the interaction is
 weaker, the spatial correlation will be weakened accordingly.
 We remark also that the above results can be generalized to the surface penetration
in any $S$-wave paired Fermi systems.

%\begin{acknowledgments}
We thank T. Nakatsukasa, K. Washiyama, K. Yabana, and K. Yoshida
for useful discussions.
This work was partly supported by
the Major State 973 Program 2013CB834400;
the National Natural Science Foundation of China under Grants No. 11335002,
No. 11005069, and No. 11175002;  the Research Fund for the
Doctoral Program of Higher Education under Grant No. 20110001110087;  and
the Grant-in-Aid for Scientific Research (No. 21340073,
No. 23540294 and No.24105008) from the Japan Society
for the Promotion of Science.

%
%\end{acknowledgments}

%\end{CJK*}
\end{document}